\documentclass[journal,onecolumn, 11pt]{IEEEtran}
\usepackage{cite}
\usepackage[table]{xcolor}
\usepackage{amsmath,amssymb,amsfonts}
\usepackage{algorithmic}
\usepackage{graphicx}
\usepackage{subcaption}
\usepackage{array}
\usepackage{arydshln}
\usepackage{balance}
\usepackage{makecell}
\usepackage{booktabs}
\usepackage[]{footmisc}
\usepackage[margin=1in]{geometry}

\ifCLASSINFOpdf
\else
\fi
\begin{document}
%
\title{Indoor Localization for IoT Using Adaptive Feature Selection: A Cascaded Machine Learning Approach}
%
%
%

\title{Indoor Localization for IoT Using Adaptive Feature Selection:\\ A Cascaded Machine Learning Approach}
\author{\IEEEauthorblockN{\textbf{Mohamed I. AlHajri}$^{*,\ddag}$, \textbf{Nazar T. Ali}$^\dag$, \textbf{Raed M. Shubair}$^\ddag$}\\
\IEEEauthorblockA{$^*$Electrical Engineering and Computer Science, Massachusetts Institute of Technology, USA \\
$^\dag$Electrical and Computer Engineering, Khalifa University, UAE \\
$^\ddag$Research Laboratory of Electronics, Massachusetts Institute of Technology, USA\\ 
Emails: $\lbrace$malhajri; rshubair$\rbrace$@mit.edu; nazar.ali@ku.ac.ae}
}
\maketitle

\begin{abstract}
Evolving Internet-of-Things (IoT) applications often require the use of sensor-based indoor tracking and positioning, for which the performance is significantly improved by identifying the type of the surrounding indoor environment. This identification is of high importance since it leads to higher localization accuracy. This paper presents a novel method based on a cascaded two-stage machine learning approach for highly-accurate and robust localization in indoor environments using adaptive selection and combination of RF features. In the proposed method, machine learning is first used to identify the type of the surrounding indoor environment. Then, in the second stage, machine learning is employed to identify the most appropriate selection and combination of RF features that yield the highest localization accuracy. Analysis is based on $k$-Nearest Neighbor ($k$-NN) machine learning algorithm applied on a real dataset generated from practical measurements of the RF signal in realistic indoor environments. Received Signal Strength, Channel Transfer Function, and Frequency Coherence Function are the primary RF features being explored and combined. Numerical investigations demonstrate that prediction based on the concatenation of primary RF features enhanced significantly as the localization accuracy improved by at least 50\% to more than 70\%.
\end{abstract}

\begin{IEEEkeywords}
Indoor Localization, IoT, Machine Learning, $k$-Nearest Neighbor, Received Signal Strength, Channel Transfer Function, Frequency Coherence Function.
\end{IEEEkeywords}

%
\IEEEpeerreviewmaketitle

\newpage

\section{Introduction}
\label{sec:introduction}
The advent of Internet-of-Things (IoT) has revolutionized the field of telecommunications opening the door for unprecedented applications such as \cite{ji_map-free_2018,benmessaoud_novel_2017,hasani_hybrid_2015,loyez_distributed_2015,hanssens_indoor_2018,goian2015fast,al-fuqaha_internet_2015,alhajri2018accurate,alhajri2015hybrid}. These applications are enabled through smart sensors that cooperate efficiently to provide the desired service. The efficient deployment of IoT-based sensors primarily depends on accurate localization and adjustment of sensor power consumption according to the radio frequency (RF) propagation channel which is dictated by the type of the surrounding
indoor environment.

The complexity of the indoor wireless channel, due to the severe multipath effect and absence of Line-of-Sight (LOS) path,  emphasizes the need for a sophisticated yet efficient approach for indoor localization. Location fingerprinting based on various machine learning algorithms has been reported in the literature using different RF features  \cite{bahl2000radar,roos2002probabilistic,kushki2007kernel,honkavirta2009comparative,lin2005performance,khanbashi_performance_2012,bevan_rf_2012,ggh1-6j32-18,al2013real,alhajri_classification_2016,xie_improved_2016,choi_robust_2017,piciarelli_visual_2016,tomic_robust_2019}. These have included Received Signal Strength (RSS), Channel Transfer Function (CTF), and Frequency Coherence Function (FCF). Developing a machine learning approach that is based on a hybrid combination of such primary RF features can greatly enhance localization accuracy, since more information will be available to the algorithm. Such an approach does not exist in the literature.

This paper presents a novel method based on a cascaded two-stage machine learning approach for highly-accurate and robust localization in indoor environments using adaptive selection and combination of RF features. In the proposed method, and based on the authors prior work \cite{alhajri2018classification,alhajri2018machine}, machine learning is first  used to identify the type of the indoor environment based on real data measurements of the RF signal in different indoor scenarios. Then, in the second stage, machine learning is employed to identify the most appropriate selection and combination of RF features that yield the highest localization accuracy. The most appropriate selection and combination of RF features of the second stage is dependent on the type of indoor environment, which is being identified from the first stage.     Analysis is based on $k$-Nearest Neighbor ($k$-NN) machine learning algorithm applied on a real dataset generated from practical measurements of the RF signal in realistic indoor environments. Received Signal Strength, Channel Transfer Function, and Frequency Coherence Function are the primary RF features being explored and combined.  

\newpage

\section{Signatures of Indoor Environment}
\subsection{Primary RF Features}
In a multipath rich indoor environment, the received signal  $Y(f)$ contains replicas of the RF transmitted signal $X(f)$. The ratio of $Y(f)$ to $X(f)$ defines the CTF, $H(f)$, which contains the multipath effects of the wireless channel. Hence, $H(f)$ can be represented as the superposition of the gains associated with the multipath components as follows \cite{nerguizian_geolocation_2004}:
 
\begin{equation}
H(f)=\sum\limits^{L}_{l=1}a_l\exp{[-j(2\pi f\tau_l-\theta_l)]}
\label{1a}
\end{equation}%
where $a_l$, $\tau_l$, and $\theta_l$ are the amplitude, delay, and phase of the $l^{th}$ multipath component; $L$ is the total number of multipath components; and $f$ is a frequency within bandwidth of operation of the channel. The analysis in this paper considers an indoor scenario with a high Signal-to-Noise Ratio (SNR)\cite{alsindi2014empirical,khanbashi2016measurements}. The CTF, $H(f)$, is considered as an RF feature that would be distinctively unique for every spatial position within the indoor environment. Under frequency selective fading, this RF signature becomes more sensitive to channel variations due to the rapid fluctuations of the gains associated with its multipath components.

The complex autocorrelation of CTF, $H(f)$, defines another channel metric known as FCF, $R(f)$, given by \cite{al2013real}:
\begin{equation}
R(f)=\int\limits^{\infty}_{-\infty}H(\hat{f})H^*(\hat{f}+f)~d\hat{f}
\label{1ca}
\end{equation}%

FCF given in (2) represents the frequency domain coherence of the radio channel. It is interpreted as the autocorrelation of the CTF, $H(f)$, due to different frequency shifts. FCF is known for its slow varying nature in the spatial domain\cite{al2013real}, which makes it a strong candidate as an RF signature.

\subsection{Hybrid RF Features}
Characterizing the type of an indoor environment by a machine learning algorithm becomes a more accurate process when the algorithm makes decisions based on concatenation of primary RF features of the same environment. In such a way, the accuracy of prediction increases due to the wealth of information available to the algorithm from the multi-dimensional dataset constructed from the concatenation of the combined primary RF features of the same environment. Hence, the approach is based on  \textit{hybrid} features. The investigations were carried out using multi-dimensional information generated from various possible combinations of all or subset of RSS, CTF, and FCF which form the hybrid features. Results, presented later in section 5, based on these hybrid features show indeed a substantial improvement in the algorithm prediction resulting in a significant enhancement in environment separability  and localization accuracy. 

\begin{figure}[h]
       \centering
        \includegraphics[width=0.65\textwidth,clip]{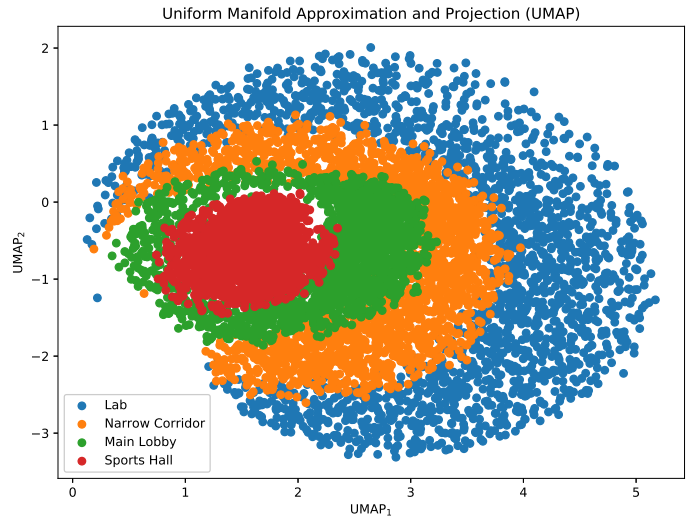}
        \caption{Using hybrid RF features in UMAP algorithm allowed different environments to be classified into distinctive clusters.}
                 \label{UMAP1}
\end{figure}

Uniform Manifold Approximation and Projection (UMAP) is a non-linear dimensionality reduction technique that is well-suited for embedding high-dimensional data for visualization in a low-dimensional space \cite{mcinnes2018umap}.  In this paper, we mapped the 5-D hybrid RF feature vector (RSS + CTF + FCF) into 2-D, with results visualized in Fig. \ref{UMAP1}. It should be noted that the hybrid RF feature vector (RSS + CTF + FCF) is 5-D since it is the concatenation of the real-valued RSS followed by the real and imaginary parts of CTF and FCF, respectively. Fig. \ref{UMAP1} indicates that the resulting 2-D mapped hybrid RF features vector of the Sports Hall (open space) is closest to that of the Main Lobby (low cluttered).  Fig. \ref{UMAP1} also indicates that the 2-D mapped hybrid RF features vector of the Narrow Corridor (medium cluttered) is surrounded by that from the Main lobby (low cluttered) and Lab (highly cluttered).  The 2-D embedding of the mapped RF features in Fig. \ref{UMAP1} shows that further insights can be gained from the UMAP visualization to confirm the advantage of using the concept of hybrid RF features. 

\newpage

\section{Dataset Generation from Real Measurements}
A dataset has been generated through a measurement campaign in which readings of the RF signal were taken in realistic indoor environments of four different types within Khalifa University campus, Sharjah, UAE. The frequency domain CTF, $H(f)$, was obtained for four indoor environments: highly cluttered (Laboratory), medium cluttered (Narrow Corridor), low cluttered (Lobby), and open space (Sports Hall). 

The measurement setup was similar to that used in \cite{alhajri2018machine} has been utilized. The dataset generated for each environment consists of 196 grid points $\times$ 10 iterations/(grid point). The dataset has been divided into 75\% training dataset and 25\% testing dataset. The dataset has been made available as open-access online so that it can be utilized for further explorations. It can be accessed online as per details in \cite{UCIML24GHZ,ggh1-6j32-18}.  

\section{Machine Learning Algorithm}
The machine learning algorithm adopted in this paper is $k$-Nearest Neighbor ($k$-NN) which is a non-parametric method used for classification \cite{altman_introduction_1992,scikit-learn}. In this algorithm, the $k$ neighbors of each hybrid RF features vector will determine the type of environment. The $k$-neighbors are associated with the shortest $k$ Euclidean distances, which are defined as follows \cite{bahl2000radar}:  
\begin{equation}
    d(\mathbf{s}_m ,\mathbf{s}_n) = \Vert \mathbf{s}_m -  \mathbf{s}_n \Vert_2 = \sqrt{\sum_{i=1}^{K} \left[s_m(i) -  s_n(i)\right]^2}
    \label{eucledian}
\end{equation}
where $K$ is the length of the RF features vector, $\mathbf{s}_m$ and $\mathbf{s}_n$ are the vectors of hybrid RF features at positions $m$ and $n$, respectively.

\begin{figure}[h]
       \centering
        \includegraphics[width=0.55\textwidth,clip]{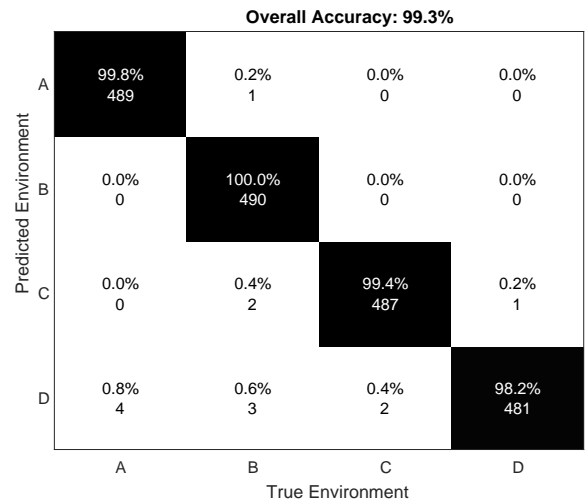}
        \caption{Confusion matrix using $k$-NN algorithm ($k$ = 1) based on hybrid RF features CTF + FCF.}
                 \label{Confusion matrix}
\end{figure}

\newpage
\section{Cascaded Machine Learning Approach}

\subsection{Machine Learning for Indoor Environment Identification}
This section is based on prior work by the authors \cite{alhajri2018machine,alhajri2018classification} where a machine learning approach was used for indoor environment classification based on real measurements of the RF signal. Extensive numerical investigations \cite{alhajri2018machine,alhajri2018classification}  showed that a machine learning algorithm using $k$-NN method, utilizing a hybrid combination of CTF and FCF, outperforms other methods such as Decision Tree \cite{quinlan1983learning} and Support Vector Machine (SVM) \cite{cortes_support-vector_1995} in identifying the type of the indoor environment with a classification accuracy of 99.3\%. The required time was found to be less than 10$\mu$s, which verifies that the adopted $k$-NN machine learning algorithm is a successful candidate for real-time IoT deployment scenarios. The accuracy of the adopted technique is represented in terms of the confusion matrix, shown in Fig. \ref{Confusion matrix}, for which the diagonal elements represent the percentage of accurate prediction for each type of environment. The off-diagonal elements of the confusion matrix represent the percentage of misclassification. The results in Fig. \ref{Confusion matrix} have been obtained for 490 test cases per environment.

\begin{table*}[h]
\centering
\caption{Comparison of localization distance error, RMSE, for different types of indoor environments using $k$-NN algorithm ($k$ = 1) based on various primary and hybrid RF features.}
\setlength{\extrarowheight}{2.6pt}
\scalebox{0.9}{
\label{mylabel23}
\begin{tabular}{c|c|c|c|c|c|c|c|c|}
\cline{2-8}
                                      & \multicolumn{3}{c|}{Primary RF Features}  &  \multicolumn{4}{c|}{Hybrid RF Features}                                            \\ \cline{2-9} 
                                      & RSS    & CTF   & FCF   & RSS + CTF & RSS + FCF & CTF + FCF & RSS + CTF + FCF & $\alpha$\\ \hline
\multicolumn{1}{|l|}{A: Lab}             & 109.19 cm & 39.75 cm & 49.56 cm & 39.99 cm     & 100.25 cm    & \textbf{39.68 cm}     & 39.9 cm & \textbf{63.7\%}          \\ \hline
\multicolumn{1}{|l|}{B: Narrow Corridor} & 105.3 cm & 30.53 cm & 45.89 cm & 31.16 cm     & 97.94 cm    & \textbf{30.49 cm}    & 30.71 cm & \textbf{71.0\%}         \\ \hline
\multicolumn{1}{|l|}{C: Lobby}           & 106.25 cm & 27.52 cm & 38.9 cm  & 27.84 cm     & 94.98 cm     & \textbf{27.18 cm}    & 27.82 cm & \textbf{74.4\%}          \\ \hline
\multicolumn{1}{|l|}{D: Sports Hall}      & 111.3 cm  & 63.33 cm & \textbf{55.2 cm}  & 63.91 cm    & 103.52 cm   & 63.32 cm     & 63.9 cm & \textbf{50.4\%}          \\ \hline
\end{tabular}}
\end{table*}

\subsection{Machine Learning for Localization Position Estimation}
 Studies in the literature indicate that a substantial gain in accuracy is achieved by improving the construction of the RF signature, rather than the use of more complex algorithms \cite{honkavirta2009comparative,wang2016csi,wang2017csi,wang2015phasefi}. As a result, in the second stage and after the type of indoor environment has been identified, a specific selection and combination of RF feature is utilized based on the identified indoor environment. We have adopted the $k$-NN algorithm again in the second stage in order to estimate the sensor position by comparing the testing RF feature to the training database of RF features, with the purpose of finding the best matching entry.   

Table \ref{mylabel23} shows the localization distance error, RMSE, using various primary and hybrid RF features. It can be seen from that the choice of the best RF signature is dependent on the type of the indoor environment. In the case of the Lab, Narrow Corridor, and Lobby, a hybrid RF feature combining CTF + FCF produced the least error RMSE, whereas for the case of the Sports Hall, a primary RF feature based on FCF has the best performance. These findings clearly indicate the importance of incorporating information on the type of the indoor environment for improving location estimation. 

In order to quantitatively assess the improvement in localization estimation when the proposed method is used, we  define $\alpha$ (last column in Table I) to represent the percentage reduction in localization distance error, RMSE, when the proposed method is used compared to the baseline method which uses RSS as the RF feature: 
\begin{equation}
\alpha = \left(\frac{\text{RMSE}_\text{RSS}-\text{RMSE}_\beta}{\text{RMSE}_\text{RSS}}\right) \times 100\%
\label{alpharatio}
\end{equation}
where $\text{RMSE}_\text{RSS}$ is the distance error due to RSS considered to be the baseline method, and $\text{RMSE}_\beta$ is the distance error due to the RF feature that produces the lowest RMSE.  It is evident from the computed values of $\alpha$ as given in the last column in Table I, that the proposed method significantly reduces the percentage error in localization by at least 50\% to more than 70\%. 

\begin{figure}[h!]
       \centering
        \includegraphics[width=0.65\textwidth,clip]{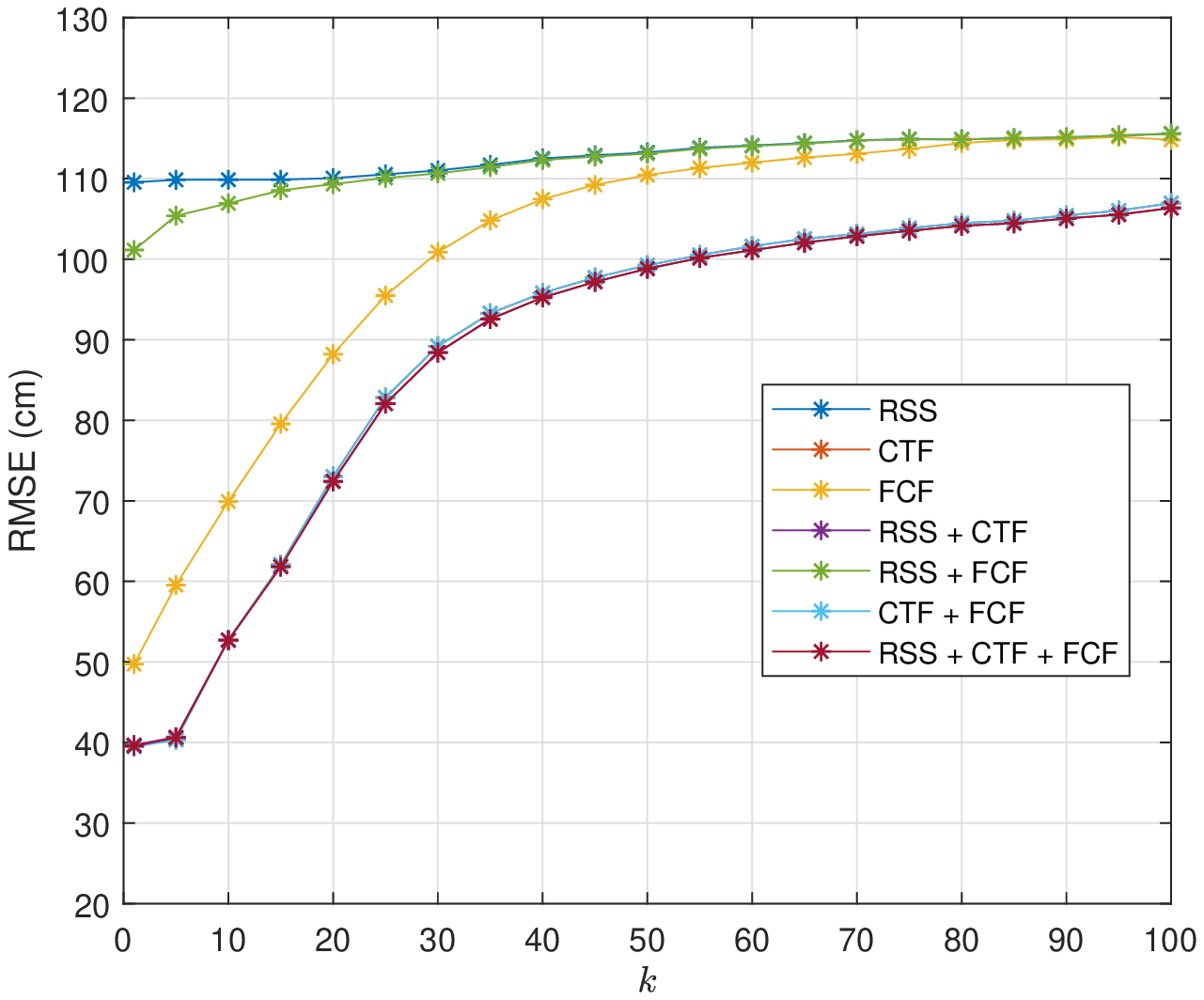}
        \caption{Comparison of RMSE using $k$-NN algorithm ($k$=1) based on various primary and hybrid RF features for Lab (highly cluttered).}
                 \label{UMAPa}
\end{figure}

\begin{figure}[h!]
       \centering
        \includegraphics[width=0.65\textwidth,clip]{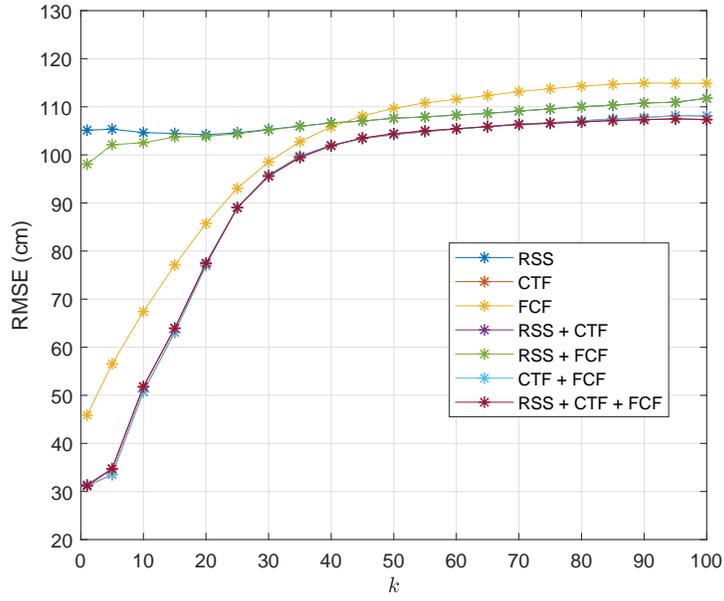}
        \caption{Comparison of RMSE using $k$-NN algorithm ($k$=1) based on various primary and hybrid RF features for Narrow Corridor (medium cluttered).}
                 \label{UMAPb}
\end{figure}

\begin{figure}[h!]
       \centering
        \includegraphics[width=0.65\textwidth,clip]{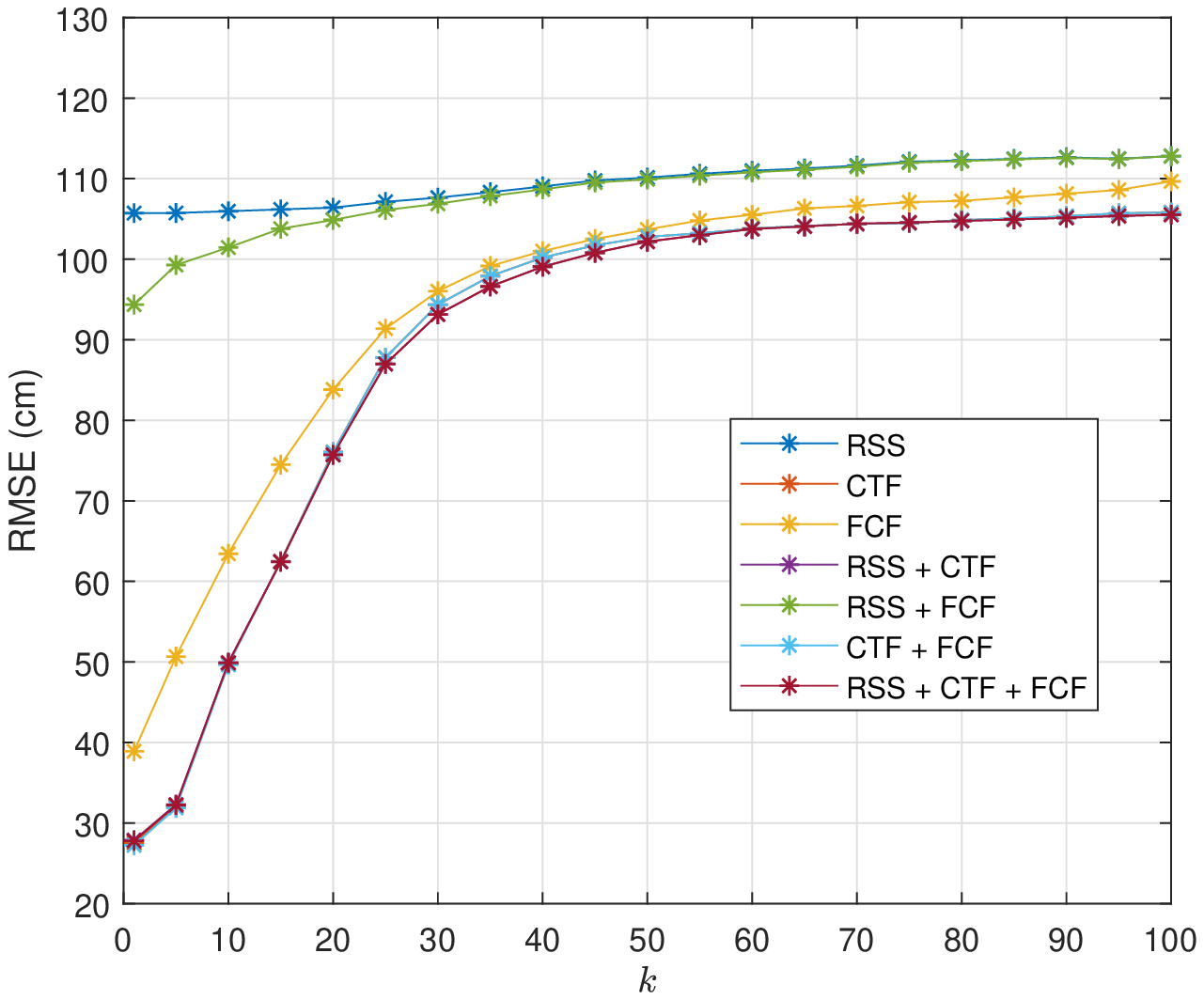}
        \caption{Comparison of RMSE using $k$-NN algorithm ($k$=1) based on various primary and hybrid RF features for Lobby (low cluttered).}
                 \label{UMAPc}
\end{figure}

We have also studied the effect of varying the number of neighbors, $k$, on the performance of different RF signatures in order to decide, for every type of environment, the best combination of primary features that yield the highest localization accuracy. The results in Fig. \ref{UMAPa} - \ref{UMAPc} indicate that increasing $k$ reduces the accuracy since the localization distance error, RMSE, increases.  This is attributed to the fact that increasing the number of neighbors, $k$, means that the algorithm attempts to find the precise location of the sensor node within a neighboring area that is large due to having many neighbors, $k$. In such case, the local information available to the algorithm reduces since prediction is made based on a larger number of neighbors, causing a deterioration in the performance of the algorithm.

\begin{figure}[h!]
       \centering
        \includegraphics[width=0.65\textwidth,clip]{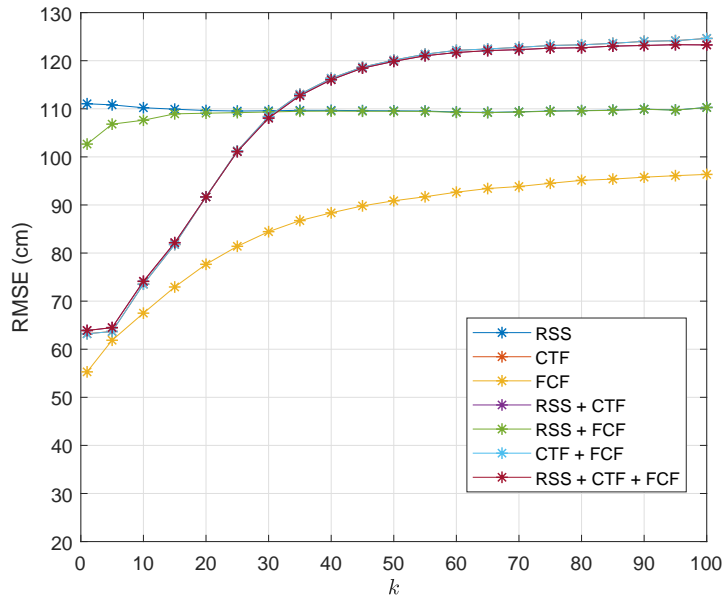}
        \caption{Comparison of RMSE using $k$-NN algorithm ($k$=1) based on various primary and hybrid RF features for Sports Hall (open space).}
                 \label{UMAPd}
\end{figure}

An interesting observation is that all RF features, both primary and hybrid, produce a steady performance when $k$ increases beyond the range $40 \leq k \leq 50$.  Another interesting observation is that primary features produce the best performance when the indoor environment has a fewer number of multipath components, as the case for indoor open space of the Sports Hall in Fig. \ref{UMAPd}, where the best performance is obtained using only FCF. On the other hand, and as evident from Figs. \ref{UMAPa} - \ref{UMAPc} which correspond to indoor environments having a large number of multipath components, the best performance is obtained using hybrid features CTF + FCF. Such interesting finding reveals the fact that using the proposed hybrid RF features concept becomes more necessary in a cluttered environment having a large number of multipath components, a scenario that is encountered in many indoor environments that involve narrow corridors or closer boundaries. 

The reduction in the localization distance error, RMSE, using the proposed method is verified by comparing it to other methods in the literature.  It is evident from Table \ref{mylabel231} that the the proposed method yields the lowest localization distance error, RMSE, compared to the other existing methods in the literature such as those by Bahl \textit{et al.} \cite{bahl2000radar}, Li \textit{et al.} \cite{li2017improved}, Alsindi \textit{et al.} \cite{khanbashi2016measurements},\cite{al2013real}.

\begin{table}[h!]
\centering
\caption{Comparison of localization distance error, RMSE, using the proposed method and other methods in the literature.}
\setlength{\extrarowheight}{1.5pt}
\scalebox{1.2}{
\label{mylabel231}
\begin{tabular}{c|c|c|c|c|c|}
\cline{2-6}
                                        &  \cite{bahl2000radar}  & \cite{li2017improved}       & \cite{khanbashi2016measurements}        & \cite{al2013real} & \begin{tabular}[c]{@{}c@{}}Proposed\end{tabular}        \\ \hline
\multicolumn{1}{|l|}{A:  Lab}              & 109.19 cm & 108.76 cm & 49.56 cm & 39.75 cm & \textbf{39.68 cm}   \\ \hline
\multicolumn{1}{|l|}{B: Narrow Corridor}  &105.3 cm  & 104.3 cm  & 45.89 cm & 30.53 cm & \textbf{30.49 cm}  \\ \hline
\multicolumn{1}{|l|}{C: Lobby}             &     106.25 cm     & 105.1 cm  & 38.9 cm    & 27.52 cm & \textbf{27.18 cm}     \\ \hline
\multicolumn{1}{|l|}{D: Sports Hall}    &  111.3 cm      & 110.9 cm      & 55.2 cm      & 63.33 cm & \textbf{55.2 cm}     \\ \hline
\multicolumn{1}{|l|}{Average Performance}    &  108.0 cm      & 107.3 cm      & 47.4 cm      & 40.3 cm & \textbf{38.1 cm}     \\ \hline
\end{tabular}}
\end{table}

\newpage

\section{Conclusion}
This paper developed a cascaded two-stage machine learning approach for highly-accurate and robust indoor localization for IoT using the concept of hybrid RF features.  In the first stage, $k$-NN algorithm was used to identify the type of indoor environment. In the second stage, $k$-NN algorithm was employed to identify the most appropriate selection and combination of RF features that yield the highest localization accuracy, based on the identified indoor environment. Investigations were carried out using a real dataset generated from practical measurements of the RF signal in realistic indoor environments. Results show that the prediction of the algorithm based on the concatenation of primary RF features enhanced significantly as the localization accuracy improved by at least 50\% to more than 70\%. The significant improvement in localization accuracy attained using the proposed two-stage machine learning method is attributed to the cascaded performance gains due to the identification of the type of environment and the concatenation of primary RF features.  

\newpage 
\bibliographystyle{IEEEtran}
\bibliography{main}

\end{document}